\documentclass[aps,prl,twocolumn,superscriptaddress]{revtex4-2}
\usepackage{graphicx}
\usepackage{hyperref}
\usepackage{amsbsy}
\usepackage{amsmath}
\usepackage{amssymb,color}
\begin{document}

\title{Strong to weak interacting-topological phase transition of bosons on a lattice}
%%%%%%%%%%%%%%%%%%%%%%%%%%%%%%%%%%%%%%%%%%%%%%%%%%%%%%%%%%%%%%%%%%%%%%%%%%%%%%%%%%%%%%%%%%%%%%%%%%%
\author{Amrita Ghosh}
\affiliation{Department of Physics, Ben-Gurion University of the Negev, Beer-Sheva 8410501, Israel}
\affiliation{Physics Division, National Center for Theoretical Sciences, Taipei 10617, Taiwan}
\author{Eytan Grosfeld}
\affiliation{Department of Physics, Ben-Gurion University of the Negev, Beer-Sheva 8410501, Israel}
%%%%%%%%%%%%%%%%%%%%%%%%%%%%%%%%%%%%%%%%%%%%%%%%%%%%%%%%%%%%%%%%%%%%%%%%%%%%%%%%%%%%%%%%%%%%%%%%%%%
\begin{abstract}
We study hard-core bosons on the honeycomb lattice subjected to anisotropic nearest-neighbor hopping along with anisotropic nearest-neighbor repulsion, using a quantum Monte Carlo technique. At half-filling, we find a transition from strong interacting-topological order to weak interacting-topological order as function of the hopping anisotropy. The strong topological phase is characterized by a finite topological entanglement entropy, while the weak topological order is identified with a non-trivial value of the bipartite entanglement entropy. Some of the order parameters and their derivatives demonstrate abrupt changes when varying the parameters controlling the lattice anistropies, thus revealing the nature of this interacting-topological phase transition.
\end{abstract}
%%%%%%%%%%%%%%%%%%%%%%%%%%%%%%%%%%%%%%%%%%%%%%%%%%%%%%%%%%%%%%%%%%%%%%%%%%%%%%%%%%%%%%%%%%%%%%%%%%%
\maketitle
\emph{Introduction.}--- In recent years, the study of topological phases in bosonic systems has become a research frontier \cite{meier2016observation,lohse2016thouless,st2017lasing,klembt2018exciton,de2019observation,jamadi2020direct,dumitrescu2022dynamical,boesl2022characterizing}. Interest in this field has been partly fueled by recent developments in optical lattice experiments, which provide a playground for realizing various phases of interacting and non-interacting bosonic systems \cite{bloch2012quantum,Atala2014,aidelsburger2015measuring,stuhl2015visualizing,goldman2016topological,tai2017microscopy,cooper2019topological}. In contrast to fermions, repulsive interactions are necessary to stabilize topological phases due to the condensation property of bosons. Interactions can also substantially enrich the topological phases compared to their non-interacting counterparts.

Interacting topologically ordered states in two dimensions (2D) can be roughly divided into two types. One type, which we will refer to as \emph{strong interacting-topological order} (SITO), is characterized by the presence of a strong topological index, which classifies the equivalence class of hamiltonians that can be deformed to each other without closing a gap. Such phases admit a finite topological entanglement entropy (TEE) \cite{levin2006detecting,kitaev2006topological}, and they can host excitations with fractional charge and anyonic exchange statistics, leading to a unique phenomenology \cite{kalmeyer1987equivalence,kitaev2006anyons,schroeter2007spin,cooper2008rapidly,wang2011fractional,nielsen2013local,greiter2014parent,gong2014emergent,bauer2014chiral,gerster2017fractional}. The other type, which we will refer to as \emph{weak interacting-topological order} (WITO), can be constructed by stacking one-dimensional (1D) chains with inter-chain hopping. Each 1D chain admits a strong topological index that relies on intra-chain interactions, while the topological order in the full 2D system is characterized by a weak topological index, being the average index of the individual chains \cite{PhysRevB.101.224201}. An intriguing question is whether a single system can be tuned between WITO and SITO, exposing the properties of this interacting-topological phase transition.

This paper reports such a strong-to-weak topological phase transition for hard-core bosons (HCBs) on the honeycomb lattice, with anisotropies in both the tunnelings and nearest-neighbor (NN) interactions. In the isotropic tunneling limit, in the presence of large enough anisotropy in the interactions, the system realizes SITO, characterized by a quantized TEE~\cite{ghosh2021chiral}. Here we demonstrate that as the anisotropy in the tunneling increases, the system transitions into a weak interacting-topological insulator at a critical value, comprising horizontal chains with weak vertical hopping. The WITO is characterized by a universal bipartite Renyi entanglement entropy (BREE) through a vertical cut, but a vanishing TEE. We explore the variations of the different order parameters characterizing the phases across this topological phase transition. The transition is marked by a jump in the slope of the edge current.
Remarkably, interactions are crucial for generating the topological order in both the weak and the strong limits.

%%%%%%%%%%%%%%%%%%%%%%%%%%%%%%%%%%%%%%%%%%%%%%%%%%%%%%%%%%%%%%%%%%%%%%%%%%%%%%%%%%%%%%%%%%%%%%%%%%%
\begin{figure}[b]
	\includegraphics[width=0.9\linewidth]{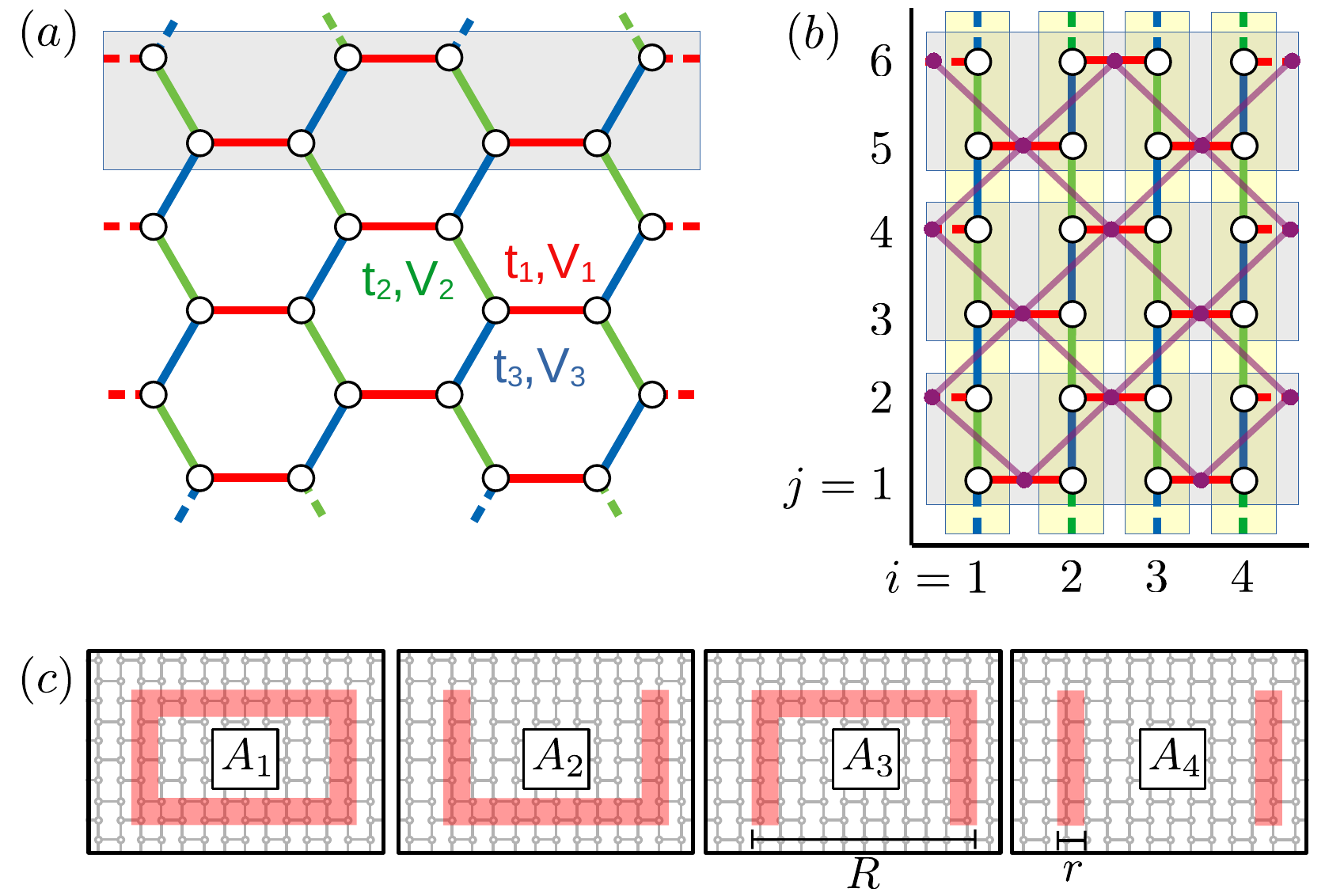}
	\caption{Pictorial description of the model on: (a), the honeycomb lattice; (b), a variant with straightened bonds. Bulk bonds (solid lines) and bonds connecting lattice sites across the boundaries (dashed lines) represent NN hopping (repulsion) of strength $t_\alpha$ ($V_\alpha$), with $\alpha=1$ (red), 2 (green) and 3 (blue).  The shaded gray regions denote the underlying 1D chains with interaction-induced dimerization. Vertical stripes (yellow) are labeled by $i$. The lattice connecting bonds of the same family $\alpha$ is marked by purple lines. (c) Subsystems $A_p$, $p=1,\ldots,4$, required for the calculation of the TEE~\cite{levin2006detecting}.}\label{lattice}
\end{figure}
%%%%%%%%%%%%%%%%%%%%%%%%%%%%%%%%%%%%%%%%%%%%%%%%%%%%%%%%%%%%%%%%%%%%%%%%%%%%%%%%%%%%%%%%%%%%%%%%%%%
\begin{figure}[t]
	\includegraphics[width=0.9\linewidth]{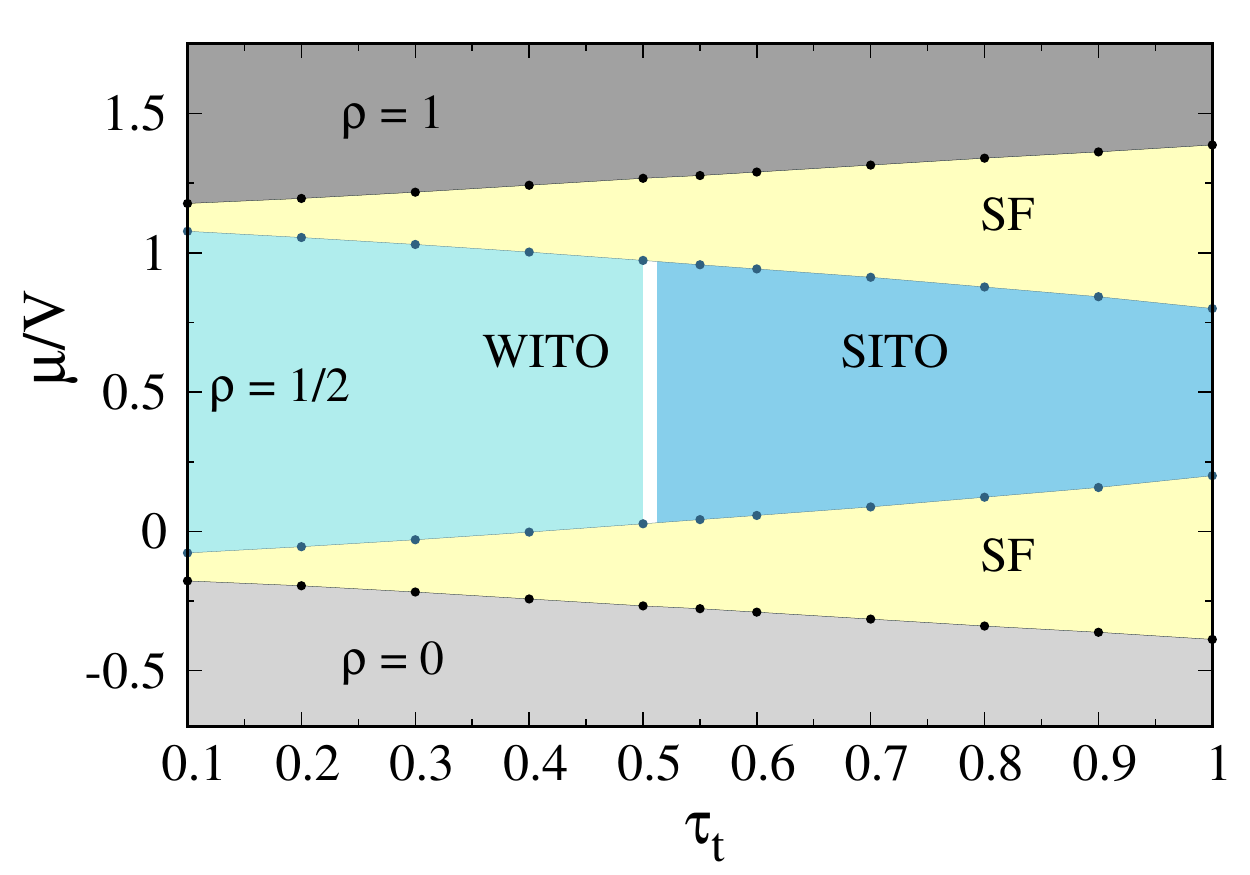}
	\caption{Complete phase diagram in terms of  $\mu/V$ and $\tau_t$ obtained for a $20\times20$ honeycomb lattice with $t=1$,$V=8$, and $\tau_V=0$. The yellow region denotes the superfluid phase, whereas the $\rho=0$, $\rho=1/2$ and $\rho=1$ represent the empty phase, dimer insulator at half-filling and Mott insulator respectively. The white area indicates the transition region between the SITO and WITO phase at half-filling.}\label{phase_diagram}
\end{figure}
%%%%%%%%%%%%%%%%%%%%%%%%%%%%%%%%%%%%%%%%%%%%%%%%%%%%%%%%%%%%%%%%%%%%%%%%%%%%%%%%%%%%%%%%%%%%%%%%%%%

\emph{The model.}--- We consider HCBs on a 2D periodic honeycomb lattice (see Fig.~\ref{lattice}), governed by the Hamiltonian
\begin{align}
 H=-\sum_{\alpha=1}^3 t_\alpha\sum_{{\langle l,m\rangle}_\alpha} \left(\hat{d}_l^\dagger \hat{d}_m+ {\rm H.c.} \right)&+\sum_{\alpha=1}^3 V_\alpha\sum_{{\langle l,m\rangle}_\alpha} \hat{n}_l \hat{n}_m\nonumber\\
 &-\sum_l \mu \hat n_l,\label{Hamiltonian}
\end{align}
where $\hat d_l^\dagger$ ($\hat d_l$) is the creation (annihilation) operator of a boson at site $l$, $\hat n_l=\hat{d}_l^\dagger \hat d_l$ is the number operator at the same site, and $\mu$ represents the chemical potential. The HCBs experience NN hopping $t_\alpha$ and NN repulsion $V_\alpha$ on bonds $\langle l,m\rangle_{\alpha}$, which belong to one of the three families $\alpha$ of parallel bonds highlighted in Fig.~\ref{lattice}. In the following we take  $\boldsymbol{t}=(t,t',t')$ and $\boldsymbol{V}=(V,V',V')$; therefore, the parameter $\tau_t=t'/t$ is a measure of the isotropy in hopping and $\tau_V=V'/V$ in the repulsive interactions. 

In the fully isotropic limit, $\tau_t,\tau_V=1$, the system realizes the t-V model~\cite{wessel2007phase}. In Ref.~\onlinecite{ghosh2021chiral}, we showed that when $\tau_V$ decreases beyond a critical value, the Hamiltonian in Eq.~\eqref{Hamiltonian} exhibits a strong interacting-topological dimer-insulator at half-filling, which admits a finite TEE $\ln(2)/2$ and chiral edge states. In this paper, we vary also the value of $\tau_t$ and construct the full phase diagram by calculating various order parameters, as well as the TEE and the (second) BREE using Stochastic Series Expansion (SSE) Quantum Monte Carlo (QMC) technique~\cite{sandvik1997finite,sandvik2010computational}. As we now describe, the additional anisotropy exposes a rich phase diagram with various topological and non-topological phases.

%%%%%%%%%%%%%%%%%%%%%%%%%%%%%%%%%%%%%%%%%%%%%%%%%%%%%%%%%%%%%%%%%%%%%%%%%%%%%%%%%%%%%%%%%%%%%%%%%%%

\emph{Order parameters and phase diagram}.---We employ the following three order parameters in order to uncover the phase diagram displayed in Fig.~\ref{phase_diagram}, as function of $\tau_t$ and $\mu/V$, at the maximally anisotropic point $\tau_V=0$. 

First, the average density of a system containing $N_s$ sites, is calculated as $\langle\hat{\rho}\rangle$ with $\hat{\rho}=\sum_l \hat{n}_l / N_s$. Here $\hat{n}_l$ is the number of HCBs (either $0$ or $1$) at site $l$ and $\langle\cdots\rangle$ represents ensemble average. Varying $\mu$, the system is found to admit three density plateaus, at $\rho=0$, $1/2$ and $1$. The plateaus at $\rho=0$ and $\rho=1$ mark the empty phase and the Mott insulator at filling-fraction $1$, respectively. 

\begin{figure}[t]
	\includegraphics[width=0.95\linewidth]{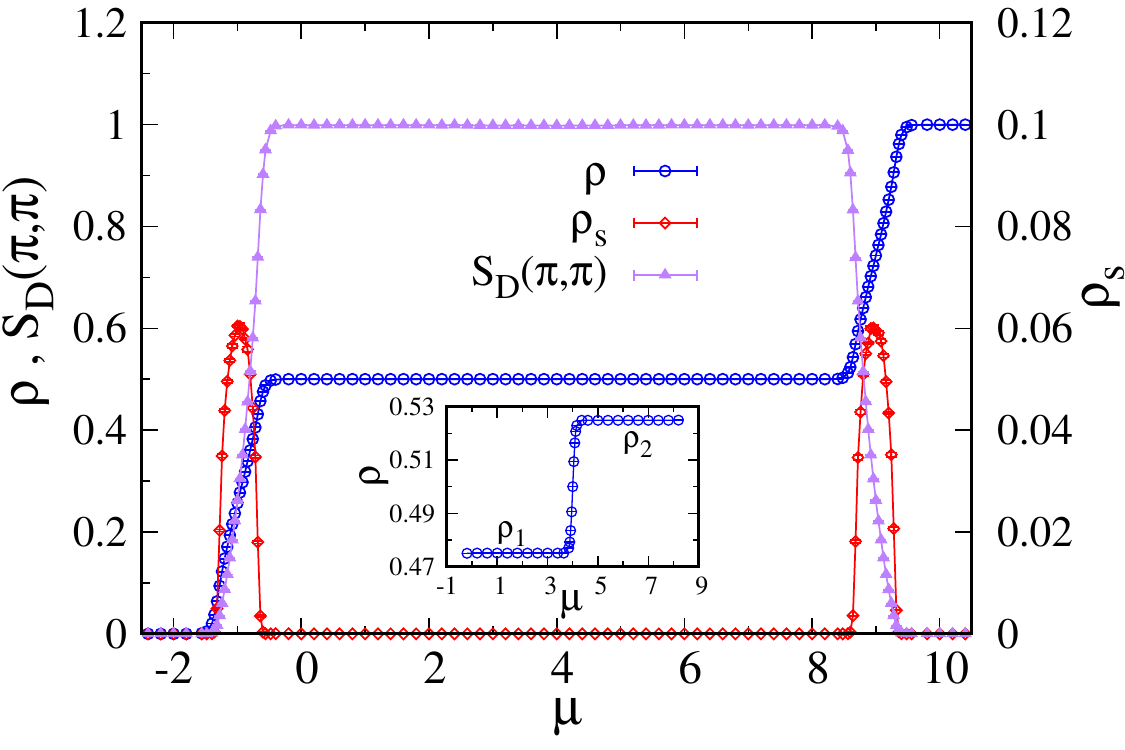}
	\caption{Plots of average density $\rho$, superfluid density $\rho_s$, and dimer structure factor $S_D(\pi,\pi)$, as function of the chemical potential $\mu$. The measurements are done on a $20\times 20$ periodic honeycomb lattice with $t=1$, $V=8$, $\tau_V=0$, $\tau_t=0.2$ and $\beta=120$. Inset figure shows the splitting of the $\rho=1/2$ plateau under open boundary conditions.}\label{order_pm_vs_mu}
\end{figure}

Next, the superfluid density is calculated as $\rho_s=\frac{1}{2}\left(\rho_s^x+\rho_s^y\right)$ where $\rho_s^a=\frac{1}{\beta}\langle \Omega_a^2\rangle$, is expressed in terms of fluctuations of winding numbers $\Omega_{a}\equiv (N_{a}^+-N_{a}^-)/L_{a}$ along the $a$-direction. Here $\beta=t/T$ is the dimensionless inverse temperature; $L_{a}$ denotes the length of the lattice along $a$-direction; and, $N_{a}^+$ ($N_{a}^-$) is the combined total number of steps the particles take in the positive (negative) $a$-direction during the evolution over an imaginary time $\beta$ to return to their original configuration of occupations.
The emergence of a density plateau at $\rho=1/2$ together with zero superfluid density (see Fig.~\ref{order_pm_vs_mu}) is a clear indicator of an incompressible insulator at half-filling. This insulating phase is surrounded by a superfluid phase (see Fig.~\ref{phase_diagram}) which separates it from the empty phase at $\rho=0$ and the Mott insulator at $\rho=1$. Noticeably the width of the insulating phase at $\rho=1/2$ increases as $\tau_t$ is decreased. This is accompanied by a decrease of the width of the surrounding superfluid region.

Finally, the dimer structure factors for the three families of bonds $\alpha$ (see Fig.~\ref{lattice}) are defined as $S^{(\alpha)}_D(\boldsymbol Q)=\sum_{bb'\in \alpha}e^{i\boldsymbol Q\cdot(\boldsymbol R_b-\boldsymbol R_{b'})}\langle \hat{D}_b \hat{D}_{b'}\rangle/N_b^2$, where $N_b$ is the number of bonds; $\boldsymbol R_b=(x_b,y_b)$ denotes the midpoint of the bond $b$ (the lattice sites of the dual lattice in Fig.~\ref{lattice}\,b); and, the dimer operator on this bond is $\hat{D}_b= \hat{d}^\dagger_{b_1} \hat{d}_{b_2}+\hat{d}^\dagger_{b_2} \hat{d}_{b_1}$ where $b_1$ and $b_2$ represent the two lattice sites attached to this bond ($b_1$ is the site either to the left or to the bottom of $b_2$). 
Due to the interactions dimers are formed only for the $\alpha=1$ family (red bonds in Fig.~\ref{lattice}), hence we focus on $S_D(\boldsymbol Q)\equiv S^{(1)}_D(\boldsymbol Q)$, while $S^{(\alpha)}_D(\boldsymbol Q)$ for $\alpha=2,3$ (blue and green bonds in Fig.~\ref{lattice}) are zero for all $\boldsymbol  Q$ values.  Since dimers are formed at all red bonds and for any pair of these bonds $(x_b-x_{b'})+(y_b-y_{b'})$ is an even number, we observe that the dimer structure factor $S_D(\pi,\pi)$ peaks with a value close to $1$ within the entire plateau at $\rho=1/2$ (see Fig.~\ref{order_pm_vs_mu}). Therefore the insulator at $\rho=1/2$ remains a dimer insulator as function of $\tau_t$ with dimers formed on every red NN bond in Fig.~\ref{lattice}. Despite its apparent uniformity, we now argue that it changes its topological nature as we decrease the value of $\tau_t$.
%%%%%%%%%%%%%%%%%%%%%%%%%%%%%%%%%%%%%%%%%%%%%%%%%%%%%%%%%%%%%%%%%%%%%%%%%%%%%%%%%%%%%%%%%%%%%%%%%%%

%%%%%%%%%%%%%%%%%%%%%%%%%%%%%%%%%%%%%%%%%%%%%%%%%%%%%%%%%%%%%%%%%%%%%%%%%%%%%%%%%%%%%%%%%%%%%%%%%%%
%%%%%%%%%%%%%%%%%%%%%%%%%%%%%%%%%%%%%%%%%%%%%%%%%%%%%%%%%%%%%%%%%%%%%%%%%%%%%%%%%%%%%%%%%%%%%%%%%%%

\emph{The topological phase transition.}--- For a strongly interacting system, such as the one we consider in this paper, the calculation of topological invariants in 2D is numerically challenging. Instead, we employ various techniques to identify the topological phase transition.

\begin{figure}[t]
	\includegraphics[width=0.9\linewidth]{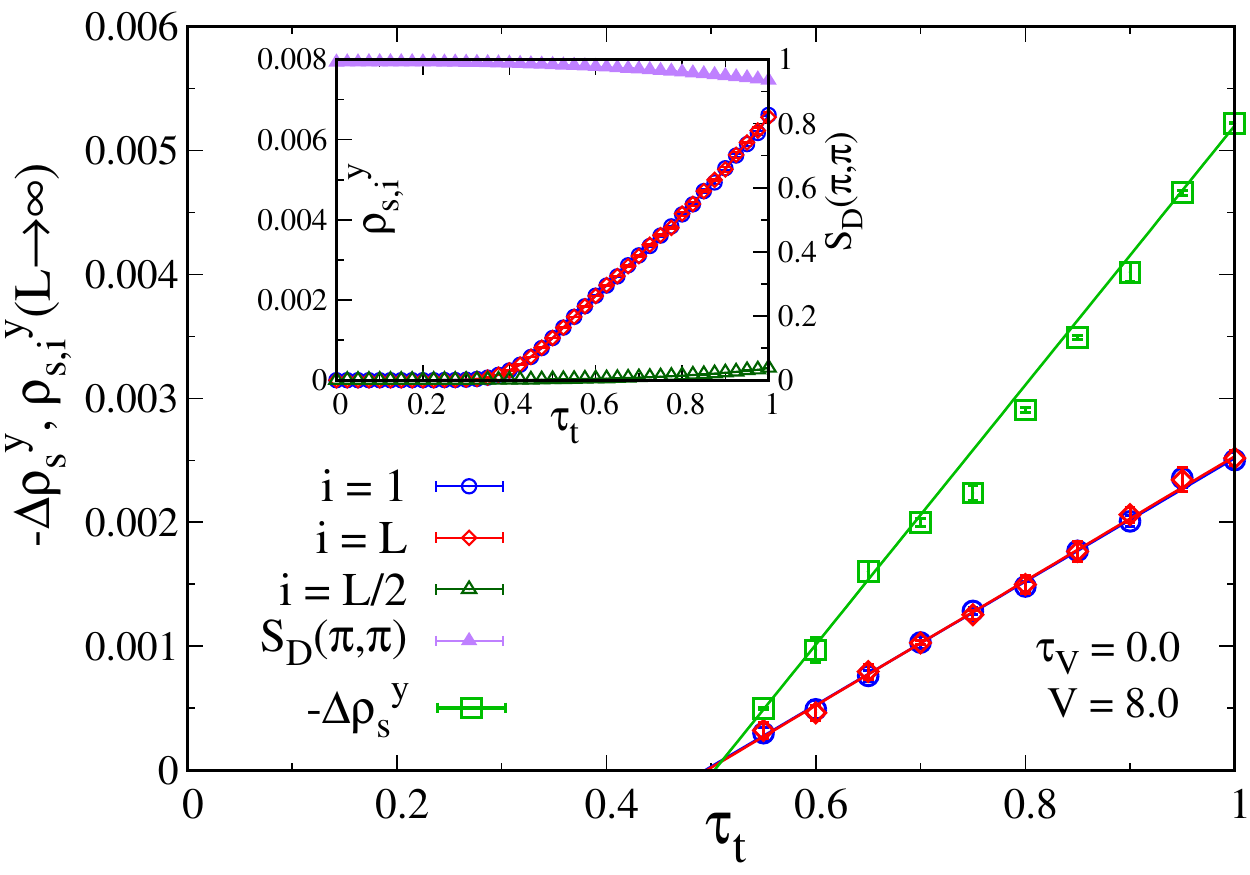}
	\caption{Variations of the superfluid density $\rho_{s,i}^y$ with the isotropy parameter of hopping $\tau_t$, measured along vertical stripes at the left edge ($i=1$), right edge ($i=L$) and in the bulk ($i=L/2$), with open boundary conditions along $x$: for a $20\times 20$ lattice (inset); and, in the thermodynamic limit, with each point extracted by finite-size scaling analysis (main panel). The dimer structure factor $S_D(\pi,\pi)$ is measured for a $20\times 20$ periodic honeycomb lattice (inset). The chirality indicator $-\Delta\rho_s^y$ is measured in the thermodynamic limit (main panel). Here $\beta=120$ and $\mu=4$.}\label{tau_vs_SF}
\end{figure}

First, since the strong topological phase at $\tau_t=1$ entails protected chiral edge states under open boundary conditions, in order to observe any transition from this phase, we study the behavior of the edge current as a function of $\tau_t$. For this purpose we define the stripe superfluid density $\rho^y_{s,i}=\frac{1}{\beta}\langle \Omega_{y,i}^2\rangle$ with the winding number projected to the $i$'th vertical stripe displayed in Fig.~\ref{lattice}\,b. \footnote{Technically, the value of $\Omega_{y,i}$ can be extracted by finding the combined total number of steps $N_{y,i}^+$ ($N_{y,i}^-$) the particles perform \emph{within the relevant stripe $i$} in the positive (negative) $y$-direction during the evolution over an imaginary time $\beta$ to return to their original configuration of occupations}. 
The inset of Fig.~\ref{tau_vs_SF} displays $\rho_{s,i}^y$ along the two edge stripes ($i=1,L$), as well as one bulk stripe ($i=L/2$), as function of $\tau_t$ for a $20\times 20$ lattice. While the bulk superfluid density remains vanishingly small throughout, the edge superfluid density decreases with $\tau_t$ and becomes nearly zero around $\tau_t=0.4$. Performing a finite-size scaling analysis %for the edge superfluid density 
at several $\tau_t$ values %upto $0.55$, 
in the main panel of Fig.~\ref{tau_vs_SF} we plot the thermodynamic-limit-extrapolated edge superfluid density as a function of $\tau_t$. It demonstrates that in the thermodynamic limit the edge superfluid density decreases with $\tau_t$ and becomes zero around $\tau_t=0.5$.
%we obtain the variation of the superfluid density for the two edge stripes in the thermodynamic limit shown .
This behavior, which manifests as an abrupt change of slope in the variation of the superfluid density as function of $\tau_t$, is our first indicator of a phase transition. Interestingly, from the inset of Fig.~\ref{tau_vs_SF}, one can see that the dimer structure factor maintains its value close to $1$ throughout the entire range of $\tau_t$. %while the structure factor remains zero.
So while the insulator at half-filling is still a dimer insulator at low $\tau_t$, its topological nature may be different. To check this possibility, next we study the entanglement properties of the system.

%%%%%%%%%%%%%%%%%%%%%%%%%%%%%%%%%%%%%%%%%%%%%%%%%%%%%%%%%%%%%%%%%%%%%%%%%%%%%%%%%%%%%%%%%%%%%%%%%%%
%%%%%%%%%%%%%%%%%%%%%%%%%%%%%%%%%%%%%%%%%%%%%%%%%%%%%%%%%%%%%%%%%%%%%%%%%%%%%%%%%%%%%%%%%%%%%%%%%%%

The $n$'th BREE in a 2D topological system follows a modified area law $S_n(A)=a\ell-q\gamma$, with $a$ a non-universal constant and $\ell$ representing the boundary length between the subsystem $A$ and its complement. The TEE is the topological component $\gamma$, which gets multiplied by the number of connected components $q$ in subsystem A. It can be extracted using Levin and Wen's construction \cite{levin2006detecting} by adding and subtracting the BREE for four different subsystems $A_p$, $p=1,\ldots,4$, as defined in Ref.~\cite{levin2006detecting},
\begin{align}
	\gamma=\lim_{r,R\to\infty}\frac{1}{2}\left[-S_n(A_1)+S_n(A_2)+S_n(A_3)-S_n(A_4)\right],
\end{align}
with $A_p$, $r$ and $R$ shown schematically in Fig.~\ref{lattice}\,c. Here $S_2(\beta)$ is accessible using QMC at finite temperature via a thermodynamic integration~\cite{PhysRevB.82.100409,isakov2011topological}. The TEE is non-zero only in a strong topological phase. Instead, it vanishes in a weak-topological and non-topological phases.

\begin{figure}[t]
	\includegraphics[width=0.9\linewidth]{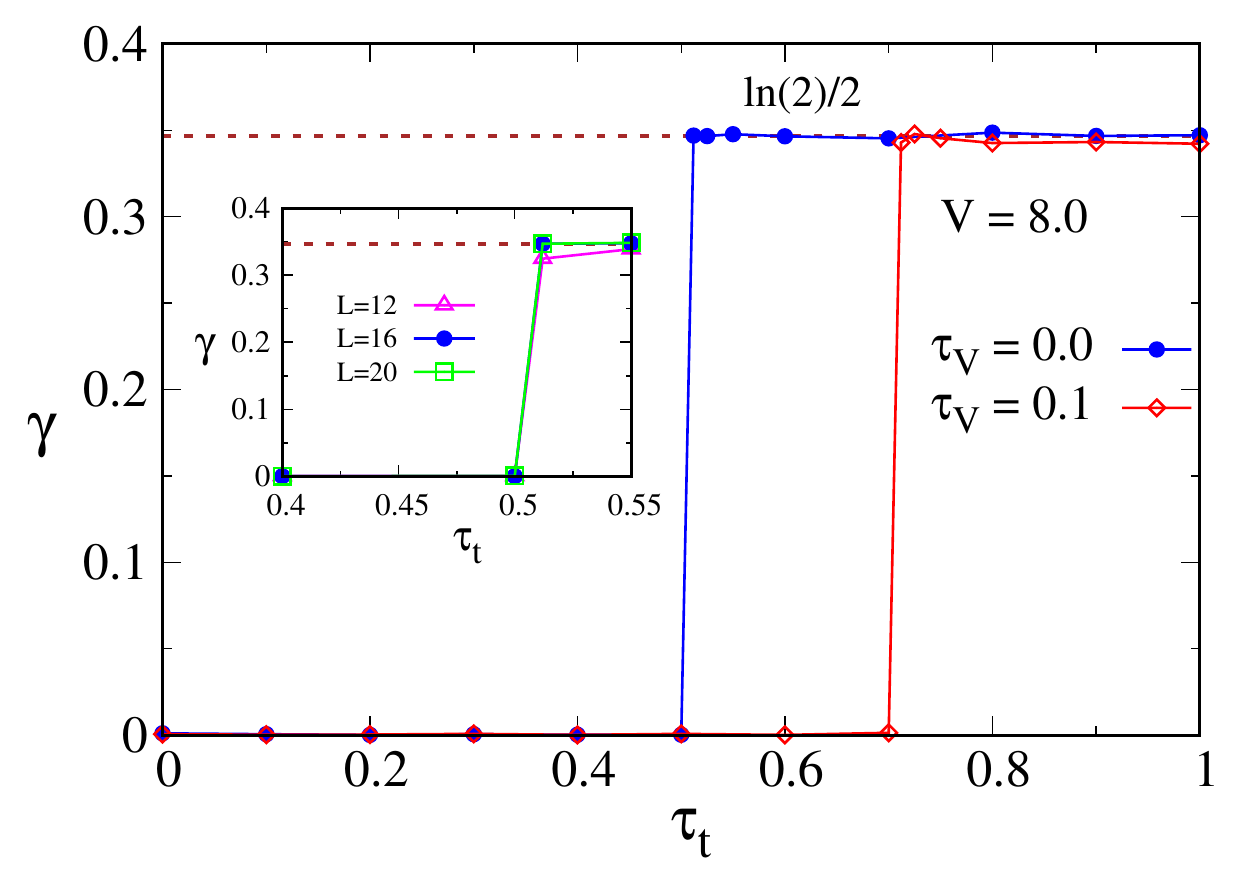}
	\caption{Variation of topological entanglement entropy $\gamma$ as a function of the isotropy parameter of hopping $\tau_t$ measured on a $16\times16$ honeycomb lattice with $t=1$, $V=8$, $\beta=1.5$, $\tau_V=0$ at $\mu=4$ ($\tau_V=0.1$ at $\mu=4.8$). Inset shows the system size dependence of the transition for $\tau_V=0$.}\label{TEE_vs_taut}
\end{figure}

To explore the possibility of a phase transition, we study the behavior of the TEE as function of $\tau_t$. The results are depicted in Fig.~\ref{TEE_vs_taut}\,: for $\tau_V=0$, as the value of $\tau_t$ is decreased from $1$ up to $\tau_t=0.512$, the TEE remains quantized at $\ln(2)/2$ revealing the existence of strong topological order. However, at $\tau_t=0.5$, the TEE suddenly drops to zero and remains zero for values below $\tau_t=0.5$, indicating a phase transition that takes place between $\tau_t=0.512$ and $\tau_t=0.5$, below which the strong phase disappears. The inset of Fig.~\ref{TEE_vs_taut} depicts this transition for three different system sizes, which demonstrates that the transition point remains unaffected by the system size. However, the position of this
transition is affected by the value of $\tau_V$. Taking instead $\tau_V=0.1$, we observe in Fig.~\ref{TEE_vs_taut} that the transition shifts to a value in between $0.712$ and $0.7$. We conclude that for larger values of $\tau_V$ the transition point shifts towards higher value of $\tau_t$, diminishing the range of $\tau_t$ where the strong topological phase is observed.

In order to probe the chirality of the edge currents in the SITO phase, we calculate the quantity $\Delta\rho_s^y=\rho_s^y-\sum_{i}\rho^y_{s,i}$. As argued in Ref.~\cite{ghosh2021chiral}, in the thermodynamic limit a measurement of 
%$\mathcal{C}=-\frac{1}{2}\Delta\rho_s^y/\rho^y_{s,E}$
$\Delta\rho_s^y\simeq-2\rho^y_{s,E}$, 
with $\rho^y_{s,E}$ being the average edge superfluidity, would indicate chirality in the system.
%when $\mathcal{C}\simeq 1$. 
Fig. \ref{tau_vs_SF} depicts the variation of $-\Delta\rho_s^y$ 
%$\mathcal{C}$ 
as a function of $\tau_t$ for an infinite lattice at $\tau_V=0$. In agreement with the TEE, it shows that indeed around $\tau_t=0.5$, the chiral nature of the edge current vanishes.
%%%%%%%%%%%%%%%%%%%%%%%%%%%%%%%%%%%%%%%%%%%%%%%%%%%%%%%%%%%%%%%%%%%%%%%%%%%%%%%%%%%%%%%%%%%%%%%%%%%
%%%%%%%%%%%%%%%%%%%%%%%%%%%%%%%%%%%%%%%%%%%%%%%%%%%%%%%%%%%%%%%%%%%%%%%%%%%%%%%%%%%%%%%%%%%%%%%%%%%

\emph{The weak interacting-topological phase.}--- The remaining question is what is the nature of the dimer insulator at $\rho=1/2$ at values of $\tau_t$ below the critical value of 0.5.

First, we explore the possibility of the existence of edge states. Under open boundary conditions along the $x$-direction, we observe that the plateau at $\rho=1/2$ splits into two equal parts (inset of Fig.~\ref{order_pm_vs_mu}) corresponding to densities $\rho_{1,2}=(L\mp 1)/(2L)$ for a $L\times L$ lattice. %($L=20$ in the inset)
This splitting indicates the existence of mid-gap edge states; the lower (upper) plateau corresponds to situation when no (all) edge sites are occupied. This suggests that the dimer insulator at $\tau_t\leq0.5$ has certain edge states associated with it, which might be topological in nature.

\begin{figure}[t]
	\includegraphics[width=0.9\linewidth]{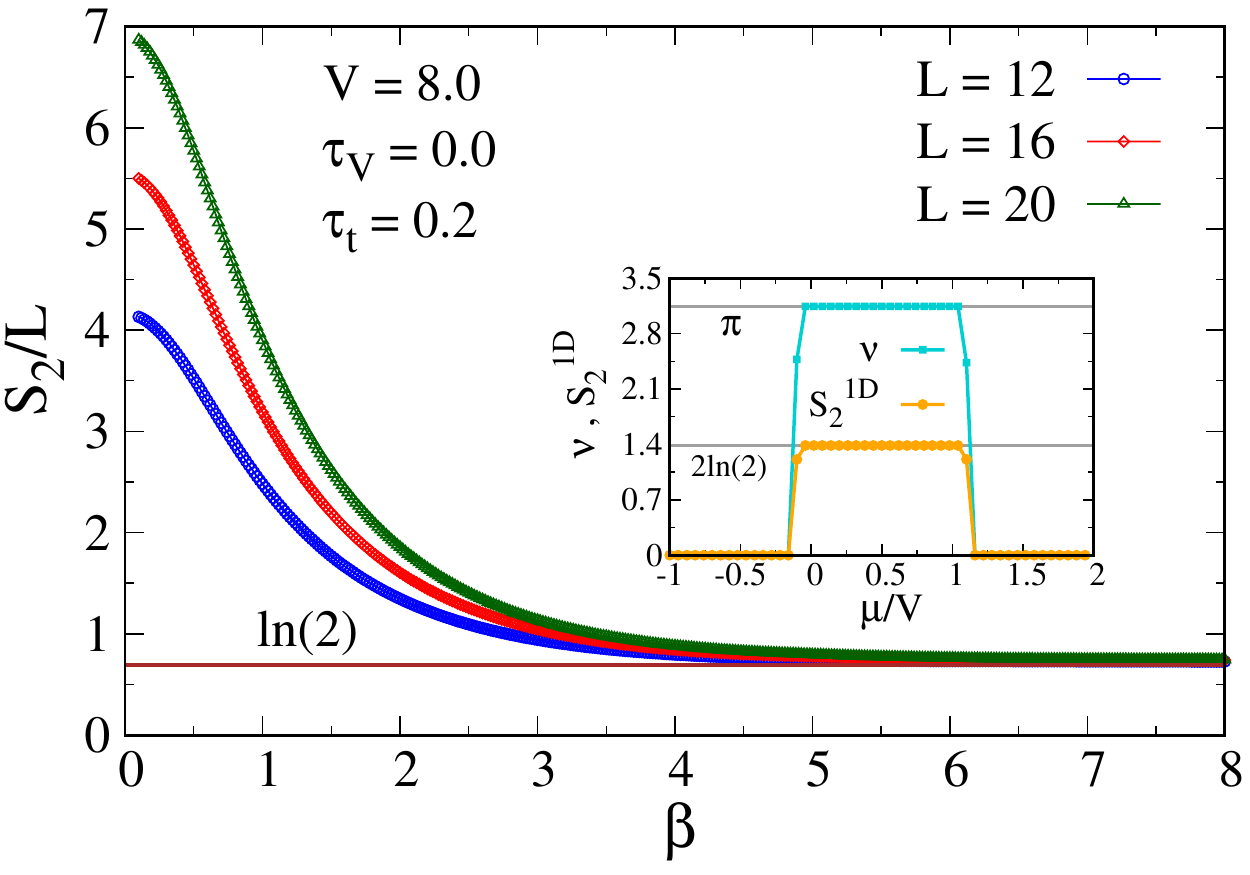}
	\caption{Bipartite Renyi entanglement entropy as a function of $\beta$ for three different system sizes, $12\times12$, $16\times16$ and $20\times20$ measured at $\mu=4$. The inset displays the 1D topological invariant and the BREE for a single 1D horizontal stripe with $8$ sites. Here $t=1$,$V=8$, $\tau_V=0$ and $\tau_t=0.2$.}\label{EE_vs_beta}
\end{figure}

Since the TEE vanishes in this phase, we turn our attention to the calculation of the BREE. We divide the periodic honeycomb lattice into two equal halves using a vertical cut and then calculate the BREE for any one of the halves. We observe $S_2/L$ for different values of inverse temperature $\beta$ to determine the constant $a$ (the proportionality factor of the area law) in the ground state of our system at a high $\beta$ value. Fig.~\ref{EE_vs_beta} compares the BREE per unit length for three different system sizes $12\times12$, $16\times16$ and $20\times20$ where the NN repulsion is fixed at $V=8$ with $\tau_V=0$ and $\tau_t=0.2$. As $\beta$ increases, the BREE approaches $\ln(2)$ for all three system sizes. 

%%%%%%%%%%%%%%%%%%%%%%%%%%%%%%%%%%%%%%%%%%%%%%%%%%%%%%%%%%%%%%%%%%%%%%%%%%%%%%%%%%%%%%%%%%%%%%%%%%%
%%%%%%%%%%%%%%%%%%%%%%%%%%%%%%%%%%%%%%%%%%%%%%%%%%%%%%%%%%%%%%%%%%%%%%%%%%%%%%%%%%%%%%%%%%%%%%%%%%%

Using exact diagonalization, we calculate the 1D strong topological invariant $\nu$ and the zero temperature BREE of a single horizontal stripe (depicted by shaded gray rectangles in Fig.~\ref{lattice}\,a,b). The topological invariant can be calculated by $\nu=\text{Im}\,\log\prod_{s=1}^{M} \langle \Omega(\phi_s)|\Omega(\phi_{s+1})\rangle/|\langle \Omega(\phi_s)|\Omega(\phi_{s+1})\rangle|$ where $|\Omega(\phi_s)\rangle$ is the ground-state at twisted boundary conditions along the $x$-direction with phase $\phi_s=\frac{2\pi}{M}s$. We find that $\nu$ becomes non-zero within the region of $\mu$ compatible with the region marked as WITO in Fig.~\ref{phase_diagram} in the small $\tau_t$ limit, while the value of the BREE becomes simultaneously $2\log 2$ (inset of Fig.~\ref{EE_vs_beta}). The natural interpretation for the 2D model is that for $\tau_t=1$ the system comprises 1D chains that are strongly connected to each other, admitting a strong 2D index as reflected in the non-zero $\gamma$. As we decrease the value of $\tau_t$, below the critical value of $\tau_t\simeq 0.5$, the insulator becomes a weak topological insulator, where the effective chains, each admitting a 1D strong index, are weakly connected, while the 2D system admits only a weak 2D index. This is reflected in the bipartite entanglement~\cite{ryu2006entanglement}: first, $\gamma$ is zero; and, second, since for a $L\times L$ honeycomb lattice there are $L/2$ such chains, $S_2$ turns out to be $L/2\times 2\ln2$, i.e., $S_2/L$ becomes $\ln2$, in perfect agreement with the QMC calculation at $\beta\to\infty$. Therefore, we finally identify the transition in Fig.~\ref{TEE_vs_taut} as a strong-to-weak  topological transition governed by the isotropy parameter of hopping $\tau_t$.

%We now argue that this is consistent with the value of the BREE for a single 1D Su-Schrieffer-Heeger (SSH) chain. This value is known to be $2\ln2$ for topologically non-trivial situation, whereas for the non-topological scenario it becomes zero. Looking at the structure of the dimer insulator at half-filling of our model, one realizes that since dimers are formed at every red NN bond in Fig.~\ref{lattice}, the underlying building block of this insulator can be effectively thought of as SSH chains (depicted by a shaded gray rectangle in Fig.~\ref{lattice}). For $\tau_t=1$, these SSH chains are strongly connected to each other. But as we decrease the value of $\tau_t$, the connections between these chains become weaker. As a result, below some critical value of $\tau_t$ (which appears to be in between $0.6$ and $0.5$ according to our QMC calculations), the insulator becomes a weak topological insulator where effective SSH chains, weakly connected to each other, are stacked along the $y$-direction of the lattice. Since for a $L\times L$ honeycomb lattice there are $L/2$ such SSH chains, $S_2$ turns out to be $L/2\times 2\ln2$, i.e., $S_2/L$ becomes $\ln2$. 

%The entanglement entropy has been argued to be in one-to-one correspondence with the Zak phase \cite{ryu2006entanglement}. 

%%%%%%%%%%%%%%%%%%%%%%%%%%%%%%%%%%%%%%%%%%%%%%%%%%%%%%%%%%%%%%%%%%%%%%%%%%%%%%%%%%%%%%%%%%%%%%%%%%%

\emph{Conclusions.}--- In this paper we studied HCBs on a honeycomb lattice subjected to anisotropic NN repulsions as well as anisotropic NN hopping. We observed that in the extreme anisotropic limit of the repulsive interactions ($\tau_V=0$), the isotropy parameter of hopping $\tau_t$ tunes a strong-to-weak interacting-topological phase transition. The phase transition is characterized by an abrupt change of the TEE when $\tau_t$ goes through a $\tau_V$-dependent critical value. In addition, the superfluid density on the edge shows a jump in its slope at this critical value. The weak phase is identified by a zero value of the TEE along with a universal value of the BREE at vanishing temperature. This is in one-to-one correspondence with the fact that weak topological phases are associated with a zero strong topological index, but a non-zero weak topological index. The weak topological phase is an interacting version of a symmetry protected topological phase, akin to the models described, e.g., in Refs.~\cite{ghosh2021weak,singh2022mirror}, but with interaction-induced dimerization. While there have been studies of the effect of interactions on topological phase transitions~\cite{ostrovsky2010interaction,li2015interacting,scheurer2015dimensional,roy2016continuous}, these phases inherit their topological properties from the non-interacting cases. Instead, in the model discussed here, both the weak and strong phases rely on interactions to manifest their topology for every value of $\tau_t$.

%%%%%%%%%%%%%%%%%%%%%%%%%%%%%%%%%%%%%%%%%%%%%%%%%%%%%%%%%%%%%%%%%%%%%%%%%%%%%%%%%%%%%%%%%%%%%%%%%%%

\begin{acknowledgments}
This research was funded by the Israel Innovation Authority under the Kamin program as part of the QuantERA project InterPol, and by the Israel Science Foundation under grant 1626/16. AG thanks the Kreitman School of Advanced Graduate Studies for support. AG would also like to thank National Science and Technology Council, the Ministry of Education (Higher Education Sprout Project NTU-111L104022), and the National Center for Theoretical Sciences of Taiwan for support towards the end of this project.
%Ministry of Science and Technology, National Center for Theoretical Sciences of Taiwan 	
\end{acknowledgments}

%%%%%%%%%%%%%%%%%%%%%%%%%%%%%%%%%%%%%%%%%%%%%%%%%%%%%%%%%%%%%%%%%%%%%%%%%%%%%%%%%%%%%%%%%%%%%%%%%%%

\bibliography{bibliography}

\end{document}